# Can Zero-Point Phenomena Truly be the Origin of Inertia?

Charles T. Ridgely
charles@ridgely.ws

A current approach to the problem of inertia suggests that the origin of the inertial properties of matter is the interaction between matter and vacuum electromagnetic zero-point radiation. Herein, it is shown that zero-point phenomena can be treated as the origin of inertia only when one chooses to ignore the mass-energy content of matter. In the absence of any physical basis for such a choice, it is concluded that zero-point-induced forces must arise in addition to the intrinsic inertial properties of ordinary matter.

## 1. Introduction

Several recent papers have proposed that the origin of the inertial properties of matter can be explained from Newton's second law of motion

$$\mathbf{f} = \frac{d}{dt}(m\mathbf{v}) \qquad (1)$$

by expressing the inertial mass $m$ in terms of the vacuum electromagnetic zero-point-field (ZPF) [1]-[5]. According to the ZPF proposal, when a body of ordinary matter is accelerated by an external force, the quarks and electrons constituting the body scatter a portion of the zero-point radiation occupying the space through which the body moves. This scattering of radiation leads to an electromagnetic drag force on the body, which according to ZPF proponents can be associated with the inertial resistance of the body [1]-[5]. The ZPF hypothesis seems compelling because it not only provides a local description of the inertial properties of matter, but also provides an electromagnetic basis for inertial and gravitational forces alike. Supporters of the ZPF hypothesis have performed a great deal of work in attempting to show that the interaction between matter and zero-point radiation gives rise to observable forces [1]-[5]. Even so, the ZPF hypothesis still seems to suffer from at least a couple of inherent conceptual difficulties.

Certainly a real zero-point radiation field must give rise some level of resistance on accelerating matter. The question is, however: What fraction of the inertial force on an accelerating object is actually induced by interaction with the ZPF? This seems to be hinged ultimately on how one chooses to view $E = mc^2$, referred to as the 'law of inertia of energy' in earlier times [6]. Those who support the ZPF hypothesis interpret the energy content of ordinary matter as entirely kinetic energy of quarks and electrons induced by interaction with zero-point radiation [1], [4]. Of course, the problem with this interpretation is that it implies that subatomic particles have no intrinsic rest mass/energy content. The present analysis employs the more traditional interpretation that $E = mc^2$ is merely a statement that all forms of energy possess inertial properties [7]-[12]. Since subatomic particles are known to possess rest mass/energy, such particles ought to exhibit intrinsic inertial properties in addition to any effects arising due to zero-point radiation.

Another problem is that the ZPF hypothesis ascribes the inertial properties of matter entirely to the inertial mass $m$ in Newton's second law of motion [1]-[5], while ignoring the underlying participation of space-time [7]-[12]. The relativistic behavior of space-time clearly forms the basis necessary for the existence of ZPF-induced forces. Since the vacuum electromagnetic ZPF is Lorentz-invariant, if space-time were absolute in the classical sense, the ZPF would remain undetectable in all systems of reference. As a result, ZPF-induced forces would not occur. Of course, space-time is not absolute, and space-time anisotropy does give rise to observable forces, as is suggested by general relativity [13]-[18]. Based on this, the objective of the present analysis is to demonstrate that resistance forces induced by the ZPF act in addition to the intrinsic inertial properties of ordinary matter.

In the next Section, a block of matter undergoing uniform acceleration is considered. While the ZPF cannot be detected in an inertial system of reference, this is not so in an accelerating system. Within an accelerating system radiation is Doppler-shifted, giving rise to an observable anisotropy in the electromagnetic mode structure of the ZPF. Observers residing on the accelerating block detect a flux of zero-point radiation passing through the block. The force density due to this flux of radiation is derived for the case of a small, particle-sized block. The resulting expression makes it straightforward to see that ZPF-induced forces arise due to the behavior of space-time.

In Section 3, the expression for the force derived in Section 2 is applied to the case in which a body accelerates uniformly through zero-point radiation. To derive the total resistance force acting on the body, the expression for the force is modified to include the intrinsic rest mass-energy possessed by the accelerating body [7]-[12]. The form of the resulting expression makes it clear that zero-point phenomena cannot be the origin of inertia. Forces induced by interaction with zero-point radiation arise in addition to the intrinsic inertial properties of matter.

## 2. The Resistance Force on Accelerating Matter due to Zero-Point Radiation

Consider a block of matter, of volume $V_0$, undergoing uniform acceleration. Observers residing in the comoving reference frame (CMRF) of the block detect a flux of Doppler-shifted zero-point radiation entering the block through the front side, which we may call wall $A$, and passing out of the block through the opposite side, which we call wall *B*. According to these observers, the radiation within the volume of the block possesses a momentum density of the form

$$\Delta \mathbf{p} = \frac{1}{c^2}(\mathbf{S}_A - \mathbf{S}_B) \qquad (2)$$

where $\mathbf{S}_A$ and $\mathbf{S}_B$ are the Poynting vectors corresponding to the flux of zero-point radiation detected at walls *A* and *B*, respectively. Expressing the Poynting vectors in terms of the energy density of the ZPF at each wall leads to

$$\mathbf{S}_A = c u_A \mathbf{n}, \quad \mathbf{S}_B = c u_B \mathbf{n} \qquad (3a, b)$$

where $u_A$ and $u_B$ are the energy densities of the ZPF at walls *A* and *B*, and $\mathbf{n}$ is a unit vector in the direction of the block's acceleration. Since the block is accelerating, zero-point radiation gains energy as it traverses the length of the block, from wall *A* to wall *B*. As a result, radiation detected at wall *B* appears blue-shifted relative to radiation detected at wall *A*. Taking this into account, the Poynting vector given by Eq. (3b) can be expressed as

$$\mathbf{S}_B = c u_A (d\tau_A / d\tau_B)^2 \mathbf{n} \qquad (4)$$

where $d\tau_A$ and $d\tau_B$ are intervals of proper time experienced by comoving observers situated at walls *A* and *B*, respectively. Using Eqs. (3a) and (4) in Eq. (2), the momentum density of zero-point radiation within the block becomes

$$\Delta \mathbf{p} = \frac{u_A}{c}\left[1 - \left(\frac{d\tau_A}{d\tau_B}\right)^2\right]\mathbf{n} \qquad (5)$$

Expressing this in terms of coordinate time in the accelerating system, and simplifying a bit, then leads to

$$\Delta \mathbf{p} = \frac{u_A}{c}\left(\frac{d\tau_A}{dt}\right)^2 \left[\left(\frac{dt}{d\tau_A}\right)^2 - \left(\frac{dt}{d\tau_B}\right)^2\right]\mathbf{n} \qquad (6)$$

where $dt$ is an interval of coordinate time in the accelerating system. Equation (6) shows that the momentum density of zero-point radiation within the block is a direct manifestation of time inhomogeneity in the accelerating system [10]-[12]. Were time absolute in the classical sense, the expression within square parentheses guarantees that the zero-point momentum density would be zero in all systems of reference.

The force density possessed by zero-point radiation passing through the block can be derived by using $\mathbf{f} = \Delta \mathbf{p}/\Delta \tau$, where $\Delta \tau$ is an interval of proper time in the CMRF of the block. Comoving observers at wall *A* can express $\Delta \tau$ in terms of the block's length, $\Delta x'$, along the $x'$–axis of the CMRF, by noticing that the time taken for a light signal to complete a round trip across the block is $\Delta \tau = 2\Delta x'/c$. Using this time interval and Eq. (6), the force density within the block can then be expressed as

$$\mathbf{f} = \frac{u_A}{2\Delta x'}\left(\frac{d\tau_A}{dt}\right)^2\left[\left(\frac{dt}{d\tau_A}\right)^2 - \left(\frac{dt}{d\tau_B}\right)^2\right]\hat{\mathbf{x}}' \qquad (7)$$

where $\hat{\mathbf{x}}'$ is a unit vector in the direction of the block's acceleration along the $x'$–axis of the CMRF. Equation (7) holds for large volumes in which $d\tau_A \neq d\tau_B$ and $\Delta x'$ assumes sizeable values. However, when the volume of the block is very small, then $d\tau_A \approx d\tau_B$ and $\Delta x'$ assumes very small values. The force density within a particle-sized volume can be expressed by taking the limit of Eq. (7) as $d\tau_B$ tends to $d\tau_A$ and $\Delta x'$ tends to zero:

$$\mathbf{f}' = \lim_{\substack{d\tau_B \to d\tau_A \\ \Delta x' \to 0}} \mathbf{f} \qquad (8)$$

Carrying out the limit and noting that $dt/d\tau_B > dt/d\tau_A$, the force density assumes the form [11]

$$\mathbf{f}' = -u_0 \frac{\partial}{\partial x'}\left[\ln\left(\frac{dt}{d\tau}\right)\right]\hat{\mathbf{x}}' \qquad (9)$$

where $u_0$ is the proper energy density of the ZPF according to observers residing in the CMRF of the particle.

Equation (9) expresses the force density of zero-point radiation passing through a particle of matter undergoing substantial acceleration along the $x'$–coordinate axis of the CMRF. When the acceleration is weak, however, the force density simplifies to [10]-[11]

$$\mathbf{f}' = -u_0 \frac{\partial}{\partial x'}\left(\frac{dt}{d\tau}\right)\hat{\mathbf{x}}' \qquad (10)$$

And when transformed from the accelerating system to flat space-time, the force density becomes [10]

$$\mathbf{f} = -u_0 \nabla (dt/d\tau) \qquad (11)$$

where it is assumed that $v \ll c$, and the prime has been dropped from $\mathbf{f}'$ for simplicity.

According to observers in flat space-time, when an external force $\mathbf{f}$ is exerted on a particle, a resistance force $\mathbf{f}_{ZPF}$ arises due to interaction between the particle and zero-point radiation. Using Eq. (11), observers in flat space-time can express this resistance force as

$$\mathbf{f}_{ZPF} = -u_0 V_0 \nabla (dt/d\tau) \qquad (12)$$

where $V_0$ is the volume of the accelerating particle. Equation (12) gives the resistance force acting on an accelerating particle due to the scattering of zero-point radiation.

According to Eq. (12), the force on the particle arises not only due to the presence of zero-point radiation, but also due to the inhomogeneity of time within the accelerating system, characterized by $dt/d\tau$ [10]-[12]. It is interesting to note that when $dt/d\tau$ assumes a constant value, the force given by Eq. (12) is then zero. This suggests that the relativistic nature of space-time is intimately involved in the generation of ZPF-induced forces.

## 3. The Total Resistance Force on a Uniformly Accelerating Body

According to the ZPF hypothesis, when a material body undergoes acceleration due to an external force, the quarks and electrons constituting the body scatter a portion of the Doppler-shifted zero-point radiation passing through the body. The energy density of this portion of radiation is [1]-[5]

$$u_{ZPF} = \int \eta(\omega) \frac{\hbar \omega^3}{2\pi^2 c^3} d\omega \qquad (13)$$

in which the fraction of zero-point radiation that actually interacts with the accelerating body is governed by the spectral function $\eta(\omega)$. The interaction between the body and zero-point radiation imparts a quantity of energy to the body. However, another form of energy that must also be taken into account is the intrinsic rest mass-energy of the accelerating body [7]-[12]. Thus, the total energy of the accelerating body must then be

$$E = E_0 + V_0 \int \eta(\omega) \frac{\hbar \omega^3}{2\pi^2 c^3} d\omega \qquad (14)$$

where $E_0$ is the rest mass-energy of the body.

In order to determine an expression for the total resistance force acting on the accelerating body, Eq. (12) must be amended to include all forms of energy possessed by the body [7]-[12]. This calls for replacing $u_0 V_0$ in Eq. (12) with the body's total energy, $E$. Carrying this out and then inserting the total energy given by Eq. (14) leads to

$$\mathbf{f} = -\left( E_0 + V_0 \int \eta(\omega) \frac{\hbar \omega^3}{2\pi^2 c^3} d\omega \right) \nabla \left( \frac{dt}{d\tau} \right) \qquad (15)$$

According to stationary observers, this is the resistance force acting on a uniformly accelerating body of rest mass-energy $E_0$ that interacts with zero-point radiation. To simplify Eq. (15), we may consider a particular body accelerating uniformly along the $x$-coordinate axis. For this case, we may put [12]

$$dt/d\tau \approx 1 + ax/c^2 \;,\; \nabla(dt/d\tau) = \mathbf{a}/c^2 \qquad (16\text{a, b})$$

wherein the acceleration vector is $\mathbf{a} = a\hat{\mathbf{x}}$. Substituting the second of these expressions into Eq. (15), and simplifying a bit, then leads to

$$\mathbf{f} = -m_0 \mathbf{a} - \left( \frac{V_0}{c^2} \int \eta(\omega) \frac{\hbar \omega^3}{2\pi^2 c^3} d\omega \right) \mathbf{a} \qquad (17)$$

in which $E_0 = m_0 c^2$ was used to simplify the first term. Equation (17) is the total resistance force exerted on a body of mass $m_0$ that is accelerating uniformly through zero-point radiation

## 4. Conclusions

Herein, it was shown that ZPF-induced forces arise solely due to space-time anisotropy within accelerating systems, and that such forces act in addition to the inertial of matter. Supporters of ZPF hypothesis, however, claim that the inertial properties of matter are due entirely to scattering of zero-point radiation by quarks and electrons comprising matter. Clearly, this makes sense only when one chooses to ignore the intrinsic energy content of matter. This is nothing more than a choice, of course; there exists no rational basis for purposely dismissing the intrinsic energy content of matter. The interaction between matter and zero-point radiation must certainly lead to forces on accelerating objects [1]-[5], but such forces must arise in addition to the intrinsic inertia of matter.